\documentstyle[proceedings]{crckapb}

\begin{opening}
\title{\bf OPTICON:  EC Optical Infrared Coordination Network for Astronomy}
\subtitle{\bf http://www.astro-opticon.org}
\author{Gerard GILMORE}
\institute{Institute of Astronomy\\
University of Cambridge, Madingley Road, \\ Cambridge  CB3 OHA, UK
\\ gil@ast.cam.ac.uk}
\end{opening}
\runningtitle{The OPTICON EC Coordination network for astronomy}
\begin{document}

\section{ABSTRACT}

OPTICON, the ICN OPTICAL INFRARED COORDINATION NETWORK FOR ASTRONOMY,
brings together for the first time the operators of all Europe's medium to
large optical-infrared telescopes, the largest corresponding data
archives, and several user representatives. The OPTICON partners work with
their communities to identify those major challenges for the future
development of European optical-infrared astronomy which require
Europe-wide collaboration.  OPTICON sponsors and coordinates developments
towards these goals, involving the entire astronomical community through
workshops and meetings targeted towards these agreed common goals of
general importance.

\section{INTRODUCTION}

OPTICON, the Co-ordination Network for optical and infrared astronomy,
is an EC funded {\sl Infrastructure Cooperation Network} under the
{\em Enhancing Access to Large Infrastructures} part of the
Framework~5 (FP5) Human Potential Programme. Such networks are funded
to bring together infrastructure operators and `typical'
users.  OPTICON brings together Europe's multinational, national and
major regional providers of astronomical infrastructures, together
with four `representative' research institutes.  The classes of
infrastructure of direct relevance to OPTICON include optical and
infrared telescopes, their instrumentation, existing medium-sized
observatory infrastructures, data archives and their relevant
communication infrastructures, and optimization of the scientific
development and exploitation of these facilities.  By identifying and
encouraging common approaches to those challenges which require
Europe-wide collaboration, the OPTICON partners work to enhance both
the quality and the quantity of access to those research
infrastructures across the whole EU community.

\section{BRIEF HISTORY}

In 1999 the EC 5th Framework Program funded the thematic network
Optical and Infrared Co-ordination Network for Astronomy (OPTICON:
HPRI-1999-40002).  This network brings together 14 partners,
representing the major astronomical funding and management
organizations within the European Union. The OPTICON network is
sponsored to facilitate co-ordination of key developmental issues in
European astronomy. 

As part of its FP5 initiative, the EC made a deliberate effort to
encourage improved coordination and collaboration in  the
development of, and access to, European-scale and internationally
competitive research infrastructures. In practise, this meant
extending the established system of 
{\sl Infrastructure Cooperation Networks}, which existed in many
branches of science which received significant EC funding support.

The role of these networks was described at the time as:

\vskip0.5cm
\noindent{\sl INFRASTRUCTURE COOPERATION NETWORKS}

{\sl The objective of this scheme is to catalyze the self-coordination and
the pooling of resources between infrastructure operators in order to
foster a culture of cooperation between them, to generate critical
mass for research into higher performance techniques, instrumentation
and technologies, to spread good practice, to promote common
protocols and interoperability, to encourage complementarity, and to 
stimulate the creation of "distributed" and "virtual" large facilities.

Participants in these networks will be operators of research
infrastructures, research teams in universities, in research centres
and in industry, representatives of users of the infrastructures, and
equipment manufacturers.  Each network will contain at least three
mutually independent legal entities which operate research
infrastructure and which come from at least three different countries
of the Member States and Associated States (one of which at least
must be a Member State) and must be co-ordinated by one of these legal
entities.

Infrastructure cooperation networks will be implemented as thematic networks.}

\vskip0.5cm

One such network, EVN-JIVE, was in place supporting radio astronomy,
and in particular Very Long Baseline Interferometry (VLBI), and its central
data processing facility at JIVE, the Joint Institute for VLBI in
Europe, based in the Netherlands. JIVE/EVN has as partners all the
facilities (including several outside the EU geographical borders)
which manage radio telescopes in the VLBI network. Many of the users
work in the same Institutes, so that user representation is provided
naturally. 

To establish a comparable scale network in the rest of astronomy,
ensuring full community participation and support, is not easy: there
are very many Institutes in Europe active in astronomy, very many
observatories, and many major data centres. In order to ensure that
OPTICON had validity from its start, the EC invited to meetings in
Brussels one or more representatives from every major
astronomy-related funding agency, as well as the PIs of every
EC-funded astronomy-related network and grant. Over 100 people were
involved in these meetings. By February 1999 clear agreement had been
reached that establishing such a network was desirable, the major
infrastructure operators in Europe had all agreed their support,
identified the issues which they wished the network addressed as a
minimum, and the present author was asked to write and coordinate the
proposal.

The proposal was submitted to the early May 1999 proposal round, and
approved. Funding for travel and workshops over four years
was provided to ensure open community-wide participation in agreed goals.
The OPTICON partners met formally for the first time in April, 2000.

Identification of the partners was, in most cases,
self-evident. Organizations which operate observatories and large data
centres are readily identifiable. There was only one existing grant
holder under the extant Access Program, and that Institute (Instituto
de Astrofisica de Canarias, manager of the European Northern
Observatory Access grant), while not yet an operator of medium sized or
large telescopes (pace GRANTECAN), or a data centre, does operate an
observatory site, the Canarian Observatories. 

One issue which arose at once, and which remains subjective, was
selection of the `representative user groups'. Clearly no such
selection has any meaning when hundreds of comparable Institutes
exist. In practise however, the role of the users among the partners
has been restricted to hosting open scientific workshops, and 
to scientific support for a single item in OPTICON's activities,
development of the science case for future Extremely Large Telescopes.
  Since this role provides no direct benefits for the user
representatives, the relevant organizations do indeed act on behalf of
the wider community. The user Institutes were in fact recommended for
inclusion, by the EU meeting, to be the EARA members, as being an available,
independently-defined, European-wide group of major research Institutes
with a tradition in international collaboration.

The European Association for Research in Astronomy (EARA), was founded
in December 1991, joining the CNRS astrophysics laboratoire, Institut
d'Astrophysique de Paris, with the astronomy departments of the
Universities of Cambridge and Leiden, in the frame of a CNRS
initiative for "Associated European Laboratories". EARA was later
extended to include the Instituto de Astrofisica de Canarias, and the
Max Planck Institut f\"{u}r Astrophysik, all five of whose members are
OPTICON partners.

\section{ORGANIZATION AND FUNDING}

The Opticon Network includes 14 formal participants and a number of
associated partners.  The formal participants include the major
European and National astronomical agencies plus representative user
institutes. The five EARA institutes are among the latter.

\begin{table}[htb]
\begin{center}
\caption{The OPTICON Partner Organizations}
\begin{tabular}{|l|l|}
\hline
{\bf Contact Individual} & {\bf Partner Organization} \\
\hline 
 OPTICON COORDINATOR ORGANISATION & \\
{Dr Paul Murdin} & {Particle Physics and Astronomy Research Council}\\
\hline
OPTICON CHAIRMAN/CONTACT & \\
{Professor Gerard Gilmore} & {The University of Cambridge/Institute of
Astronomy}\\  
\hline
{Professor Francoise Genova} & {Universite Louis Pasteur - Strasbourg CDS}\\
\hline
{Professor Piero Benvenuti} & {European Space Agency - Space Sciences Division}\\
\hline
{Professor Alvio Renzini} & {European Southern Observatory}\\
\hline
{Professor Alain Omont} & {CNRS/Institut d'Astrophysique de Paris}\\
\hline
{Dr Genevieve Debouzy} & {Institut National des Sciences de l'Univers} \\
 & {du Centre National de la Recherche Scientifique}\\ 
\hline
{Professor Francisco Sanchez} & {Instituto de Astrofisica de Canarias}\\
\hline
{Professor Marcello Rodono} & {Consorzio Nazionale per L'Astronomia
e L'Astrofisica}\\
\hline
{Professor George Miley} & {Universiteit Leiden/Astronomy Department}\\
\hline
{Professor Simon White} & {Max Planck Institut f\"{u}r Astrophysik}\\
\hline
{Professor Hans-Walter Rix} & {Max-Planck-Institut f\"{u}r Astronomie}\\
\hline
{Professor Dr Tim De Zeeuw} & {Netherlands Research School for
Astronomy (NOVA)}\\ 
\hline
{Dr Leo Takalo} & {Nordic Optical Telescope Scientific Association}\\
\hline
\end{tabular}
\end{center}
\end{table}

The contract partners are independent national agencies or research
institutes, and multi-national organizations. Each partner is
represented by national research and funding directors, research group
directors or the equivalent.

Overall Network coordination is provided largely by the science
coordinator assisted by the OPTICON Administrator and secretary.  This
small team ensures adequate information and administrative support for
the working groups' and partners' meetings, enhances reliable and
effective communications across the network, maintains the webpage and
enhances Europe-wide information about the OPTICON activities.

The OPTICON management board meets twice a year with the inaugural
meeting being held at the National Maritime Museum, Greenwich, London
in April 2000.  The schedule of future meetings is given in our diary
available on the Opticon webpage www.astro-opticon.org

The Network operates a two-level structure.  This means that the
contract partners meet to specify timely areas of common interest and
opportunity for development and cooperation.  These areas of mutual
interest are developed and quantified where appropriate by specialist
working groups, chaired by a partner, bringing together relevant
complementary expertise and users from the whole European astronomical
community, explicitely including countries and Institutes 
not explicitly included in the present partners.

Each working group is led by a delegated partner, who is responsible
for specific management, and for reporting to the network overall. In
practise, there is a considerable degree of overlap in membership of
the working groups, so that informal communications are
excellent. Regular communications are utilized on a daily basis with
more permanent, and public, information being provided on a series of
web sites. There is a dedicated network home page
http://www.astro-opticon.org. Each working group also has its own home
page as follows:\\
www.roe.ac.uk/atc/elt/workshop/index.html\\
www.ip.de/Euro3D/\\
www.stecf.org/$~$jwalsh/OPTICON3D\\
ecf.hq.eso.org/astrovirtel/\\
www.roe.ac.uk/ifa/surveys\\

\subsection{PROFILES OF THE OPTICON PARTNERS}

A summary profile of the fourteen Opticon partners is provided below:

1) PARTICLE PHYSICS AND ASTRONOMY RESEARCH COUNCIL

The Particle Physics and Astronomy Research Council (PPARC). PPARC
funds UK research, education and public understanding in its four
broad areas of science - particle physics, astronomy, cosmology and
space science.  PPARC has three scientific sites: the UK Astronomy
Technology Centre (UKATC) in Edinburgh, the Isaac Newton Group of
telescopes (ING) in La Palma and the Joint Astronomy Centre (JAC) in
Hawaii.  http://www.pparc.ac.uk

2) INSTITUTE OF ASTRONOMY, UNIVERSITY OF CAMBRIDGE

The Institute of Astronomy is a department of the University of
Cambridge.  It is the largest centre for astronomical research in the
UK and is among the oldest scientific research departments of the
University.  The 120 staff, students and visitors are drawn from many
countries making it an international research centre dedicated to
teaching and research in many areas of observational and theoretical
astronomy.  http://www.ast.cam.ac.uk

3) CENTRE DE DONN\'{E}ES ASTRONOMIQUES DE STRASBOURG (CDS)

The Centre de Donn\'{e}es astronomiques de Strasbourg (CDS) is a data
centre dedicated to the collection and worldwide distribution of
astronomical data and related information.  It is located at the
Strasbourg Astronomical Observatory, France.

The CDS develops reference databases and tools, widely used by the
astronomy community, and collaborates actively with other data
centres, ground and space-based observatories and electronic journals
to build links between distributed on-line resources.
http://www.astro.u-strasbg.fr/obs-E.HTML

4) EUROPEAN SPACE AGENCY - SPACE SCIENCES DIVISION

ESA, the European Space Agency, provides a vision of Europe's future
in space, and of the benefits for people on the ground that satellites
can supply. It also develops the strategies needed to fulfil the
vision, through collaborative projects in space science and
technology.

Most OPTICON-related activity is organized through the ESA/NASA
Space Telescope-European Coordinating Facility

The Science Archive Facility has over twelve years of experience in
the management and development of astronomical archives and databases.
Throughout this period the Archive has pursued a steady and effective
collaboration with the CADC (Canadian Astronomy Data Centre) and has
implemented a number of innovative features.  These additions have all
proven so useful and popular that they have been adopted by other
archive sites and have become part of a set of 'minimum requirements'
for modern astronomical archive systems.  http://www.esa.int
http://www.stecf.org/astrovirtel/ http://www.stecf.org/

5) EUROPEAN SOUTHERN OBSERVATORY

ESO, the European Southern Observatory, was created in 1962 to
establish and operate an astronomical observatory in the southern
hemisphere, equipped with powerful instruments, with the aim of
furthering and organizing collaboration in astronomy

It is supported by eight countries: Belgium, Denmark, France, Germany,
Italy, the Netherlands, Sweden and Switzerland; the United Kingdom is
to join ESO in 2002.  Portugal has a cooperation Agreement with ESO,
leading to future membership.

ESO operates at two sites.  It operates the La Silla observatory in
the Atacama desert, 600 km north of Santiago de Chile, at 2,400 m
altitude, where fourteen optical telescopes with diameters up to 3.6 m
and a 15-m submillimetre radio telescope (SEST) are now in operation.
In addition, ESO is in the process of building the Very Large
Telescope (VLT) on Paranal, a 2,600 m high mountain approximately 130
km south of Antofagasta, in the driest part of the Atacama desert.
The VLT consists of four 8.2-meter and several 1.8-meter telescopes.
These telescopes can also be used in combination as a giant
interferometer (VLTI).  "First Light" of the first 8.2-meter telescope
(UT1) occurred in May 1998.  UT1 became available on a regular basis
for astronomical observations from April 1999.  Over 1000 proposals
are made each year for the use of the ESO telescopes.

The ESO Headquarters are located in Garching, near Munich, Germany.
This is the scientific, technical and administrative centre of ESO
where technical development programmes are carried out to provide the
La Silla and Paranal observatories with the most advanced instruments.
There are also extensive astronomical data facilities.  In Europe ESO
employs about 200 international Staff members, Fellows and Associates;
in Chile about 50 and, in addition, about 130 local Staff members.
http://www.eso.org/

6) INSTITUT D'ASTROPHYSIQUE DE PARIS

The Institut d'Astrophysique de Paris (IAP) is a laboratory of the
Centre National de la Recherché Scientifique (CNRS).  Founded in
1938 with the development of modern astrophysics and the foundation of
CNRS, IAP has a long history of prominent activity in observation and
theory and of international collaboration.  Its present activity
focuses on extragalactic astronomy and cosmology, including stellar
populations and star formation in galaxies, and specific aspects of
stellar physics.

The IAP hosts data reduction centers for several major international
experiments, including the infrared survey of the southern sky
(DENIS), French participation in the NASA ultra-violet mission FUSE,
TERAPIX, the data reduction center for the 1º x 1º
MEGACAM camera to be installed on the Canada France Hawaii Telescope
(CFHT), and participation in the data analysis of the ESA PLANCK
cosmic microwave background space mission.
http://www.iap.fr/accueil.html

7) INSTITUT NATIONAL DES SCIENCES DE L'UNIVERS (CNRS)
 
The Institut National des Sciences de l'Univers (INSU) is part of the
Centre National de la Recherche Scientifique (CNRS), the main
scientific public research organization in France.  INSU has the
responsibility in three scientific areas: ocean-atmosphere, earth
science and astrophysics.  Its 128 research and service units (most of
them associated with Universities) represent a total staff of 5608
individuals.  INSU is also strongly involved in large international
collaboration and participate to the funding and operation of some of
the major large ground-based infrastructure facilities.
http://www.insu.cnrs-dir.fr/

8) CONSORZIO NAZIONALE PER L'ASTRONOMIA E L'ASTROFISICA (CNAA) and
ISTITUTO NAZIONALE DI ASTROFISICA (INAF)

The Italian "National Consortium for Astronomy and Astrophysics"
(CNAA) is based in Rome and was established in 1996 by the 12 Italian
Astronomical Observatories of the Ministry of University and Research
(Arcetri-Florence, Bologna, Brera-Merate, Cagliari,
Capodimonte-Naples, Collurania-Teramo, Catania, Padua, Palermo, Rome,
Turin, Trieste) as a temporary Institution devoted to the promotion
and management of national projects, primarily the newly completed
"Telescopio Nazionale Galileo" at the Roque de Los Muchachos
Observatory (La Palma, Canary Islands), and of coordinated research
activities carried out at different institutions in Italy. The CNAA
has also served as a forum for debating questions related to the
national science policy in Astronomy.

The Italian Observatories and the CNAA are now being restructured into
a single national institution, the "Istituto Nazionale di Astrofisica"
(INAF), which is based in Rome and will take over all legal and
management responsibilities starting from mid 2001.
http://w3c.ct.astro.it/cnaa

9) INSTITUTO DE ASTROFISICA DE CANARIAS

The Instituto de Astrofisica de Canarias (IAC) is a highly
internationalized research centre and comprises:

The Instituto de Astrofisica, which constitutes the headquarters,
based in La Laguna (Tenerife, Spain); the Observatorio del Teide, in
Izaña (Tenerife); and the Observatorio del Roque de los Muchachos,
in Garafia (La Palma).

The IAC's headquarters is located on the campus of the University of
La Laguna, where it has become a meeting point for the international
astronomical community, a centre for research, technological
development and training of researchers, engineers and technicians.
The Grantecan 10 m telescope is the major undergoing technological
project. IAC is also an active promoter of science education.
http://www.iac.es

European Northern Observatory

The IAC Observatories (the Observatorio del Teide, on Tenerife, and
the Observatorio del Roque de los Muchachos, on La Palma), together
with the research facilities from 18 different countries, constitute
the European Northern Observatory (ENO), Europe's organization for
Astronomy in the North.  http://www.iac.es/eno/

10) LEIDEN OBSERVATORY

The Institute of Astronomy at Leiden University, the Sterrewacht
Leiden (Leiden) Observatory), has a long tradition and an
internationally acknowledged reputation for education and research in
astronomy.

The Institute offers all the facilities needed to participate in top
level research. The research interests of the Sterrewacht Leiden cover
many aspects of modern astronomy, ranging from stars and the
interstellar medium, to galaxies and cosmology.
http://www.strw.leidenuniv.nl/

11) MAX-PLANCK-INSTITUT F\"{U}R ASTROPHYSIK

The Max-Planck-Institut f\"{u}r Astrophysik is one of more than 70
autonomous research institutes within the Max-Planck-Society.  These
institutes are primarily devoted to fundamental research.  Most of
them carry out work in several distinct areas, each led by a senior
scientist who is a "Scientific Member" of the Max-Planck Society.

Research at MPA is devoted to a broad range of topics in theoretical
astrophysics.  Major concentrations of interest lie in the areas of
stellar evolution, stellar atmospheres, supernova physics,
astrophysical fluid dynamics, high energy astrophysics, galaxy
structure and evolution, the large-scale structure of the Universe,
and cosmology.  http://www.mpa-garching.mpg.de

12) MAX-PLANCK-INSTITUT F\"{U}R ASTRONOMIE

The Max-Planck-Institut f\"{u}r Astronomie (MPIA) in Heidelberg
operates the Calar Alto Observatory as well as conducting research in
different areas of astronomy and astrophysics.  It is one of the
Max-Planck-Institutes in Germany within the Max-Planck-Gesellschaft
(MPG) and one of the five astronomically orientated institutes in
Heidelberg http://www.mpia-hd.mpg.de http://www.caha.es

13) NETHERLANDS RESEARCH SCHOOL FOR ASTRONOMY (NOVA)

The Netherlands Research School for Astronomy, Nederlandse
Onderzoekschool voor Astronomie or NOVA's scientific program is based
on three multiply-connected inter-university networks.  It is built
around key researchers with international reputations, who lead groups
in their respective institutions (at the Universities of Amsterdam,
Groningen, Leiden and Utrecht), and who already have ongoing
collaborations.

Nova's mission is two-fold: i) to carry out front-line astronomical
research in the Netherlands and ii) to train young astronomers at the
highest international level.  http://www.strw.leidenuniv.nl/nova

14) NORDIC OPTICAL TELESCOPE SCIENTIFIC ASSOCIATION

The Nordic Optical Telescope (NOT) Scientific Association (NOTSA) was
founded in 1984 to construct and operate a Nordic telescope for
observations at optical and infrared wavelengths.

The associates members are: Statens naturvidenskabelige
forskningsr\r{a}d (Denmark); Suomen Akatemia (Finland); H\'{a}sk\'{o}i \'{I}slands (Iceland); Norges forskningsr\r{a}d (Norway); Naturvetenskapliga
forskningsr\r{a}det (Sweden).

The executive bodies of NOTSA are the NOT Council and the Directorate.
http://www.astro.utu.fi/ http://www.not.iac.es

\section{OPTICON ACTIVITY: WHAT and HOW?}

Since the start of the network, six major aspects of European
astronomical research in which there are clear benefits from
international cooperation, and where inadequate cooperation currently
exists, have been identified.  These are:

\noindent Activity 1: EU Elite Fellowship Program \\
Activity 2: The Astrophysical Virtual Observatory \\
Activity 3: Improved Coordination on Common Infrastructures \\
Activity 4: The Future of Medium-sized Observatories in the enlarged
EU \\
Activity 5: The Science Case for Extremely Large Telescopes \\
Activity 6: Joint Activities with the radio astronomy ICN (JIVE).

Working Groups, with full representation across the whole EU astronomy
community, have been established to implement these common objectives,
with substantial progress being made.

\subsection{Activity 1: EU Elite Fellowship Program}

The working group responsible for Elite Fellowships operates
to ensure that the best European fellowships on offer are of comparable
status and duration to those on offer in the US and 
through some European National Programmes. The goal is to make
European astrophysics as attractive a career option for the most
talented young scientists as options which are available in other
communities.

This scheme should enhance the production of excellent science within
Europe and help identify research leaders of the future.

\subsubsection{OVERVIEW OF PROGRESS}

A proposal for six, three-year postdoctoral
fellowships, namely the J H Oort Fellowships, had been submitted
to the EU Marie Curie scheme, but these had not been immediately
supported.  EU feedback strongly supported the scientific goals of the
proposal, but implied that the application was not suitable for the
programme to which it had been submitted.  

The failure of present EU structures to provide internationally
competitive fellowship and career opportunities for the most able
astrophysicists was identified, with the specific limitations in
current schemes being successfully localized.

Following a meeting between
the Director General of ESO and the head of DG-XII, the working group
chair had put together a new generic proposal for an *elite*
fellowship scheme which was hoped could be considered as an
*Accompanying Measure* in Framework VI.  The scheme could initially be
run as a pilot in two or three disciplines, including astrophysics.

The EU had said that it did not have the resource or facility to
manage such a scheme and would seek to allocate this responsibility to
a suitable Agency if the scheme was supported, though it was not clear
who this might be for astrophysics.  Consequently, a proposal to
investigate possible management structures in several disciplines was
submitted to the Call for Accompanying Measures, approved and funded,
and is underway.

Efforts continue to create and implement a scheme whereby
EU-funded Europe-wide fellowships for the most able astrophysicists
will be competitive with US opportunities.

\subsection{Activity 2: The Astrophysical Virtual Observatory}

The OPTICON partners agreed to coordinate their efforts towards the
realization of an Astrophysical Virtual Observatory for all European
astronomy.  An Astrophysical Virtual Observatory would allow all
European astronomers to partake in, and utilize, the technological
advances of the future internet (GRID) initiatives that have already
been recognized by the EC as critical to the development of the
European Research Area.  Similar efforts are under way in the US, in
response to an NSF decadal report on astronomy, and in other subjects.

\subsubsection{First Step}

The ASTROVIRTEL Project, supported by the European Commission and
managed by the ST-ECF on behalf of ESA and ESO, was the first stage in
the fruition of the AVO.  ASTROVIRTEL was aimed at enhancing the
scientific return of the ESO/ST-ECF Archive and offers to European
users the opportunity to exploit it as a virtual telescope, retrieving
and analyzing large quantities of data with the assistance of the
Archive staff.

ASTROVIRTEL is primarily concerned with implementation of
science-driven query tools spanning multiple extant data Bases, and
means to label the scientific integrity of dBases.  The approach taken
consists of building from specific astronomer led queries starting
with a few high quality and well understood dBases.  At present this
includes the HST and ESO/VLT archives followed by the rest of the ESA
mission archives.
        
A first call for proposals was announced in mid-2000, with 11
proposals received and 5 selected for further assessment and
implementation.

The advantages of the ASTROVIRTEL approach are that: the "scientific
interoperability" of different archives will be enhanced on the basis
of specific scientific requirements as contained in the approved
Proposals, the "mining tools" and the procedures for the management
and analysis of the retrieved data sets.  These will then become part
of the Archive and offered to the community.  See
ecf.hq.eso.org/astrovirtel/

It is envisaged that ASTROVIRTEL will naturally evolve into a part of
the larger AVO. In the meantime ASTROVIRTEL, with 3 years funding, is
providing an essential learning experience.

\subsubsection{Second Step}

With the background of ASTROVIRTEL, European-wide efforts are now in
place by the working group responsible for implementing an AVO.  They
are specifically preparing for the following:

1)A complete science case and set of science requirements;\\
2)A demonstration of interoperability using a small set of existing
archives  with varying degrees of VO-readiness;\\
3)An assessment of GRID technologies for astronomy including
prototyping, testbeds and the development and assessment of
scalable storage and processing facilities;\\
4)Implementation of active links to similar international initiatives
(e.g. 	NVO in the US) to prepare for the possibility of global VO
activities.

OPTICON established three working groups to investigate practical
implementation of these goals, and definition and implementation of
the Astrophysical Virtual Observatory: one to focus on the scientific
utilization of archives; one on the interoperability of archives; and
one on the necessary IT infrastructure for the exploitation of an
ever-increasing astronomical data flood. A meeting of the OPTICON
partner organizations in Strasbourg in October 2000 made explicit
recommendations to these working groups to prepare, by early 2001,
proposals to the 5th Framework RTD program for developments leading to
the Astrophysical Virtual Observatory, in such a way as to benefit the
entire EC-wide astronomical community.

Six key organizations were identified as members of the AVO Phase A
proposal in order to meet the requirements of the RTD program and the
aims outlined.  The UK ASTROGRID consortium was an existing
collaboration that was seeking e-Science funds from the UK government
to deploy GRID technologies for several astronomical programs. The
joining of the ASTROGRID consortium into the AVO proposal was a major
step in order to form an important unification of the European VO
effort and to optimize the return on available funds, together with
forming a unified interface to international efforts.

The RTD proposal was submitted in February 2001 and identified a 6.2
million euro work program over three years consisting of 718 man
months of development, testing and deployment, 1 million euros in
hardware and 100,000 euros in travel expenses. The immediate goals
have been achieved. The RTD proposal has been approved.

\noindent{An illustration usign the OPTICON Working Group on INTEROPERABILITY }

Among the tasks of the OPTICON network are to ensure improved
efficiency of access to and enhanced exploitation of ground and space
observations, together with the development of virtual access to large
data archives. One key element for increasing scientific access to
multi-wavelength, heterogeneous data is interoperability of data
archives and information services.  This allows scientists to retrieve
the data of interest for their research among the large variety of
possible information sources and be able to formulate queries to these
distributed on-line resources. On the service provider side, metadata
describing the service contents have to be implemented, and data
exchange mechanisms have to be defined and used to allow the
implementation of links between services and the integration of data
of different origins in common user interfaces.

This analysis was presented at the first general meeting of the
OPTICON network at Greenwich in April 2000, where it was agreed that a
Working Group to tackle these questions should be created. The
Interoperability Working Group aims at studying cost effective tools
and standards for improving access to and data exchange from data
archives and information services.  One important specification was to
keep to a minimum the additional workload on data providers. A
pragmatic bottom-up approach will be used, with email discussions,
targeted meetings to define and promote basic standards and generic
tools, short technical visits if necessary, and eventually prototype
implementation in some cases. Working Group members are managers of
European public databases and archives proposed by the OPTICON
collaboration.

The Interoperability Working Group's goals were presented at two major
international meetings: Virtual Observatories of the future (Caltech,
June 2000), and Mining the sky (Munich, July-August 2000), where
numerous contacts and discussions took place with potential
participants and international partners (USA, Canada). The list of
participants was further discussed after the second OPTICON general
meeting with the OPTICON collaboration members.  Exchanges of
information took place with the proposed members, to explain the
Working Group's goals, to acquire confirmation of their willingness to
participate, and identify a first set of information to be distributed
and of subjects to be discussed. A Web page is in preparation and a
meeting is foreseen in the coming months, with presentations of
problems and possible solutions by the Working Group members, together
with a few round-table discussions on specific topics of general
interests.

A targeted meeting was held in Strasbourg with the ECF-ESO AstroVirtel
managers in December 2000, to discuss the usage of common tools taking
into account their scientific requirements.

The importance of early partnership with other communities was
recognized from the beginning with contacts being immediately taken
with the European Radio Network and an OPTICON/EVN discussion
organized during the International Astronomical Union General Assembly
in Manchester.  The "Astronomy Information Network" was presented at
an EVN meeting in Madrid in November 2000. The radio network nominated
a representative to participate in the Working Group activities and to
diffuse the information in the radio community.  Data archive managers
from Australia, Canada and USA were invited to participate in the
Working Group activities and have fully contributed.

From these meetings and more generally to present the "Astronomy
Information Network" at which to discuss generic tools at the first
AstroGRID meeting in Belfast, January 2001, a coherent
Interoperability work program was thus established for the AVO
proposal.

Joint EU and US meetings have identified several key coordination
points and milestones for the future. A regular series of open
international and Europe-wide workshops, conferences, and scientific
meetings are scheduled, with OPTICON sponsorship.  The committees also
submitted a joint proposal for an IAU Symposium on VO Science to be
held in conjunction with the IAU General Assembly in Sydney 2003.

For the future, OPTICON will continue to coordinate EU-wide
development of the Virtual Observatory and implement the RTD aims.

A Review Paper `OPTICON and the Virtual Observatory' \\
(http://xxx.soton.ac.uk/multi astro-ph/0011464) \\ is available further
describing these activities.

\subsection{Activity 3: Improved Coordination on Common Infrastructures}

\subsubsection{ASTRO-WISE: OPTICON Working Group on Wide Field Imaging}

The aim of this programme is to provide a European astronomical survey
system, facilitating astronomical research, data reduction, and data
mining based on the new generation of wide-field sky survey
cameras. By joining the efforts of several National data centres
established in support of these cameras and of the ESO, the programme
establishes, through common standards, a European wide shared
computing infrastructure. The huge, many Terabyte, wide field imaging
data volumes call for a coordinated effort: the programme coordinates
the development of software tools and will support the derivation of
survey system products, such as Public Survey results, calibrated
images and catalogues of astronomical objects. These products will be
used for astronomical research, made available to archive facilities
to be addressed by parallel activities such as AVO, and are crucial
for the exploitation of the new very large telescopes.

Following a successful OPTICON meeting on Survey systems in
Edinburgh, the National data centers involved in wide-field imaging in
the Netherlands (NOVA), Italy (Capodimonte), France (TeraPix) and
Germany, together with ESO and the UK-VISTA community have taken the
initiative to prepare for a joint effort.
A full account of the talks and program of
the workshop can be found at the workshop web site: \\
http://www.roe.ac.uk/ifa/surveys.

A new consortium has been founded with all partners being prepared to
contribute significantly by providing both hardware and human
resources for a new European-wide-field-imaging initiative.  An RTD
proposal has been prepared and submitted, seeking funding for this
international collaboration. This proposal has been approved and
supported by the EU.

This remarkably swift development after the OPTICON meeting marks the
common needs and the appreciation of partners' expertise in the
consortium.  Several meetings between individuals from the data
centers have taken place and exchange of personnel is planned.  All
short-term intentions have been realized, most importantly building a
new Europe-wide collaboration.

Long term plans include the implementation of the RTD proposal goals,
together with continuation of common work towards agreed common goals.

\subsubsection{EURO-3D: OPTICON 3D-SPECTROSCOPY WORKING GROUP}

One of  the crucial ways in which European astronomy has acted in coordinating Europe-wide community has been identified by the OPTICON partners.  This has been in the development of common software tools to address data challenges common to major instrumental developments.  Following recommendation, the OPTICON partners considered the case for 3-Dimensional spectroscopic developments in European astronomy. The partners concluded 3-Dimensional spectroscopy as one of the most technologically challenging developments in optical-infrared astronomy at present, yet is one in which the scientific returns are immense.  It is one in which the European scientific community holds a significant and currently world-leading role. It has also been recognized that an essential requirement for European excellence in this technologically challenging field is improved coordination in development of the common infrastructure tools.

In response to this agreed priority need, and to meet the EC recommendation for a coordinated Europe-wide response, OPTICON established a Working Group, with the following remit:

To bring together representatives of all the European groups working in 3D spectroscopy; to share experience, software and expertise; to enhance common working methods; and to consider ways in which to apply for EC funding; to support developments of clear common benefit to the whole European astronomical community. 

 The working group accepted the OPTICON remit, and agreed to develop a proposal to the EC Research, Training Networks programme (RTN). In addition, the instrumental, software and future plans of all the groups were reviewed, and a critical item for progress agreed.

The aim of the RTN proposal, called Euro3D, is to coordinate and underpin the many potentially complementary activities underway in Europe, concentrating on providing software while training young researchers to scientifically exploit the many 3D spectroscopy instruments which are coming available on large telescopes.  However, the data from these instruments is large and complex and expertise in the community to exploit the scientific potential is not yet sufficiently
widespread.

The working group had met twice during 2000 to discuss and review the instrumental, software and future plans of all the groups.  Two open and widely advertised meetings were also held in Garching in December 2000 and in Potsdam in February 2001.

\subsubsection{Activity 4: The Future of Medium-sized Observatories in
 the enlarged EU} 

The existing medium sized (2-4metre telescope aperture on good
mountain sites) observatories have an
enormous potential for improved international cooperation, with
particular opportunities in enhanced training for the young and for
scientists in Central Europe. Additionally, considerable scientific
benefits to the whole European scientific community, together with
financial benefits to the national operating agencies, can follow from
improved coordination of operational facilities, instrumentation, and
procedures.

A working group has been established to achieve these training and
common operational aims.

Two working group meetings brought together, for the first time, the
operators and observatory directors of every 2-4m telescope in which
an EU country has a major financial partnership.  These historic
occasions led immediately to an appreciation of common requirements,
opportunities and challenges.

Facilities already in existence cover a wide range of science and
training applications, but there is little
co-ordination with respect to operation or development.  It has also
been noted that access to some facilities was already open to the entire
international community, but no financial support was available for
observers to reach the telescopes.

Four sub-groups were established to consider different areas of
possible co-operation and collaboration.  One of the aims of these
groups is to set out the principles for proposals which could be taken
to the wider community and funding agencies.

These groups have prepared a working document, which is now being used
as the basis of discussions between telescope operators, national
funding agencies, and extant user communities. When agreed with all
these communities, a joint proposal to the EU FP6 Access to Large
Infrastructures, together with related training and PHARE programmes,
will be developed.  Various bi-national and multi-national cooperative
arrangements have already been stimulated by these meetings.

An extremely ambitious programme, bringing together for the first time
all of Europe national telescope operators, has succeeded admirably.

It is proposed to develop, in detail, methods to enhance the
scientific and research training roles of extant 2-4m telescopes; to
implement bi-national and multi-national coordination of operations
and developments, and to propose to the EU FP6 programme a Europe-wide
training and research capability.

\subsubsection{Activity 5: The Science Case for Extremely Large Telescopes}

An immediate goal of this working group is to develop the science case
for future large telescopes, as that would form the basis for specific
technological developments.  An ancillary goal was to bring together
the European astronomy community to support an agreed future program
of major infrastructure developments, aimed at putting Europe at the
head of the world.

This science case will do the following:

i) define the technological studies and developments which are
necessary to build the telescope; 
ii) form the basis for future proposals for national and EU
funding support for development and construction of a
world-leading facility.

A major international workshop was held in Edinburgh, September 2000,
resulting in the 58 participants identifying and outlining key
scientific challenges which enhances the case for future technologies.

The material was assembled into a web-based "skeleton science case",
including technical background and performance comparisons between
space-based and ground-based facilities.  The science sessions
(planets and stars, stars and galaxies, galaxies and cosmology) were
summarized by the session chairs, and other contributions from
participants were included or linked. A software performance simulator
is under development, while the whole web-based information package forms
the basis for further development at the next planned workshop in the
series during Summer 2001.

The current text is available at www.astro-opticon.org/ELT.html

The early intention, to initiate development of the science case for
future extremely large telescopes, has been admirably achieved.  A
draft science case exists, based on full international participation,
which this will be further developed in the near future.  A pleasing
outcome which exceeded intention was the very high degree of
international interest and involvement in the planning and
implementation of next generation facilities.

\subsection{Activity 6: Common Activities involving all of astronomy}

Multinational organizations, such as the EU, and national funding
agencies, expect research communities to agree their priorities
internally. Competing proposals to national/international agencies
from inside a sub-discipline are mutually destructive. Conversely,
where several subsdisciplines can benefit from a similar
infrastructure investemnt, the case for that investment is
strengthened. An topical example in investement in high-bandwidth
communications infrastructure (the internet, GRID, and their
successors), where all science will benefit.

Coordinated approaches to funding agencies and strategy forums for
major projects are thus both necessary and desirable. There is at
present no natural forum in Europe to coordinate such
approaches. Thus, joint efforts by OPTICON and JIVE/EVN are underway,
to establish relevant communications. This will be initiated with
public meetings at the Joint European Astronomy Meetings.

\section{PUBLICITY, and Public Awareness}

A challenge for any new organization, especially for one involving
funding agencies from many countries, is to ensure that the wider
community is both fully involved and fully informed of activities and
opportunities. 

For OPTICON, a conscious decision was made that the first major effort 
to disseminate results of OPTICON's activities
had to await those activities. That is, wide advertising would await
some positive results. This rather non-commercial approach has been
followed.

The first major successes of OPTICON are now in place. 
\begin{itemize}
\item{The Astrophysical Virtual Observatory developments and initial funding have
been obtained}.
\item {Coordinated developments of common infrastructures have
been agreed, and funded.}
\item {Substantial development work towards an Elite
Fellowship programme is funded, and underway.}
\item {The Science Case for a
large Telescope is under multi-national development.}
\item {Europe's operators
of existing telescopes are meeting and working together.} 
\end{itemize}

All these successes have been achieved
under the sponsorship of OPTICON. All have been achieved in the first
year of activity.
Now is the time to address wider questions, and inform the wider community.
This is the next challenge for OPTICON.

\end{document}